\documentclass[namedreferences]{solarphysics}
\usepackage[optionalrh,solaromanenum]{spr-sola-addons} % For Solar Physics
\usepackage{graphicx}                    % For eps figures, newer & more powerfull
\usepackage{color}                       % For color text: \color command
\usepackage{url}                         % For breaking URLs easily trough lines
\usepackage{hyperref}

\chardef\us=`\_
\begin{document}

\begin{article}

\begin{opening}

\title{Can a Fast-mode EUV Wave Generate a Stationary Front?}
%%%%%%%%%%%%%%%%%%%%%%%%%%%%%%%%%%%%%%%%%%%%%%%%%%%
%% Authors Names
\author[addressref={aff1,aff2},corref,email={chenpf@nju.edu.cn}]{\inits{P. F.}\fnm{P. F.}~\lnm{Chen}}%\sep
\author[addressref={aff1,aff2}]{\inits{C.}\fnm{C.}~\lnm{Fang}}%\sep
\author[addressref={aff3}]{\inits{R.}\fnm{R.}~\lnm{Chandra}}%\sep
\author[addressref={aff4}]{\inits{A. K.}\fnm{A. K.}~\lnm{Srivastava}}%\sep

\address[id=aff1]{School of Astronomy \& Space Science, Nanjing University, Nanjing 210023, China}
\address[id=aff2]{Key Lab for Modern Astronomy and Astrophysics (Nanjing
	University), Ministry of Education, Nanjing 210023, China}
\address[id=aff3]{Department of Physics, DSB Campus, Kumaun University,
	Nainital 263001, India}
\address[id=aff4]{Department of Physics, Indian Institute of Technology (BHU),
	Varanasi 221005, India}
%%%%%%%%%%%%%%%%%%%%%%%%%%%%%%%%%%%%%%%%%%%%%%%%%%%
%% Runningheads
\runningauthor{P. F. Chen \textit{et al.}}
\runningtitle{Can a Fast-mode EUV Wave Generate a Stationary Front?}
%%%%%%%%%%%%%%%%%%%%%%%%%%%%%%%%%%%%%%%%%%%%%%%%%%%
\begin{abstract}
The discovery of stationary ``EIT waves" about 16 years ago posed a big challenge
to the then favorite fast-mode wave model for coronal ``EIT waves". It encouraged
the proposing of various non-wave models, and played an important role in 
approaching the recent converging viewpoint, {\it i.e.} there are two types of 
EUV waves. However, it was recently discovered that a stationary wave front can 
also be generated when a fast-mode wave passes through a magnetic 
quasi-separatrix layer (QSL). In this paper, we perform a magnetohydrodynamic 
(MHD) numerical simulation of the interaction between a fast-mode wave and a 
magnetic QSL, and a stationary wave front is reproduced. The analysis of the 
numerical results indicates that near the plasma beta $\sim 1$ layer in front of 
the magnetic QSL, part of the fast-mode wave is converted to a slow-mode MHD 
wave, which is then trapped inside the magnetic loops, forming a stationary wave 
front. Our research implies that we have to be cautious in identifying the nature
of a wave since there may be mode conversion during the propagation of the waves 
driven by solar eruptions.
\end{abstract}

%%%%%%%%%%%%%%%%%%%%%%%%%%%%%%%%%%%%%%%%%%%%%%%%%%%
%% Keywords
%
\end{opening}
%%%%%%%%%%%%%%%%%%%%%%%%%%%%%%%%%%%%%%%%%%%%%%%%%%%

\section{Introduction}\label{sec1}

One of the most intriguing phenomena discovered by the {\it EUV Imaging 
Telescope} \citep[EIT,][]{dela95} aboard the Solar and Heliospheric Observatory
(SOHO) spacecraft is the so-called ``EIT waves", which are named after the
observing telescope \citep{mose97, thom98, thom99}. ``EIT waves" are bright 
fronts visible in various EUV wavelengths \citep{will99, long08, kuma13}, such 
as 171 \AA\ \citep[formation temperature of 0.63 MK,][]{will99}, 193 \AA\ 
\citep[formation temperature 1.2 MK,][]{thom98}, and 211 \AA\ \citep[formation 
temperature 2 MK,][]{kuma13}, and 284 \AA\ \citep[formation temperature 2.25 
MK,][]{zhuk04}. They commence following coronal mass ejections (CMEs)/flare 
eruptions. It has been verified that they are more related to CMEs, rather than
solar flares \citep{bies02, chen06a, chen11}.

When ``EIT waves" were discovered, they were initially explained to be 
fast-mode waves (or shock waves) driven by CME/flare eruptions \citep{thom98, 
wang00, wu01, ofma02, vrsn02, vero08}, and they were thought to be the coronal 
counterparts of chromospheric Moreton waves \citep{thom98, grec14, grec15}. 
However, the primary drawback of the fast-mode wave model is that the 
velocities of the ``EIT waves" are typically $\sim$3 times smaller than those 
of the coronal fast-mode shock waves that are inferred from type II radio 
bursts \citep{klas00} or from chromospheric Moreton waves \citep{zhan11}. 

More importantly, soon after ``EIT waves" were discovered, \citet{dela99} and 
\citet{dela00} revealed that the wave fronts in several ``EIT wave" events are 
stationary, and the stationary ``EIT wave" front was found to be cospatial with
a magnetic quasi-separatrix layer (QSL), across which magnetic field lines 
change connectivity rapidly. Since it is generally thought that fast-mode wave 
would travel across the magnetic QSL, the discovery of stationary ``EIT wave" 
fronts led \citet{dela99} and \citet{dela00} to doubt the fast-mode wave model 
for ``EIT waves". They proposed that ``EIT waves" are related magnetic 
reconfiguration. 

Encouraged by their questioning on the fast-mode wave model, several
other models have been proposed, {\it e.g.} the magnetic fieldline stretching 
model \citep{chen02, chen05b}, successive reconnection model \citep{attr07a, 
attr07b, vand08, cohe09, cohe10}, the slow-mode wave model \citep{will07, 
wang09}, and the current shell model \citep{dela08}. Taking the magnetic 
fieldline stretching model for an example, the ``EIT waves" discovered by 
\citet{thom98} are believed to be generated by the successive stretching of 
the closed magnetic field lines pushed by an erupting flux rope below. In 
particular, with magnetohydrodynamic (MHD) numerical simulations, 
\citet{chen05b} confirmed that ``EIT waves" would stop at magnetic QSLs. The
reason is straightforward, {\it i.e.} on the other side of the magnetic QSL, 
the magnetic field lines belong to another flux system, which cannot be pushed
to stretch up by the erupting flux rope in the source active region. 

This magnetic fieldline stretching model also predicts that there should be a
fast-mode wave ahead of the ``EIT wave", which was missed by the EIT telescope
because of its low cadence. After the Solar Dynamics Observatory (SDO) was
launched in 2010, its high-cadence observations frequently revealed the
co-existence of a fast-moving EUV wave and a slowly-moving EUV wave in an
individual event \citep{chenwu11, asai12, cheng12, shen13, whit13, xue13, 
zong15}. The two wave features are also reproduced in 3D MHD simulations 
\citep{down12}. According to the magnetic fieldline stretching model 
\citep{chen02, chen05b}, the fast-moving EUV waves are the fast-mode wave/shock 
wave driven by the CME eruption, whereas the slowly-moving EUV waves correspond 
to the ``EIT waves" discovered by \citet{thom98}. It is noted in passing that 
\citet{nitt13} focused on the fastest moving EUV wave and ignored any 
slowly-moving waves behind in each event of their sample. Therefore, in our 
understanding, most of the EUV waves in their paper correspond to the fast-mode 
wave/shock wave, rather than the classical diffuse ``EIT waves".

From the above, it is fair to say that the discovery of the stationary ``EIT 
wave" front in \citet{dela99} and \citet{dela00} played a vital role in 
challenging the fast-mode wave mode for ``EIT waves" and helped colleagues 
approach a converging viewpoint that there are both wave and non-wave 
components in EUV wave events \citep{chen12, liu14, warm15}. 

\begin{figure}    %%%%%%%%%%%%%%%%%% FIGURE 1
   \centerline{\includegraphics[width=0.7\textwidth,clip=]{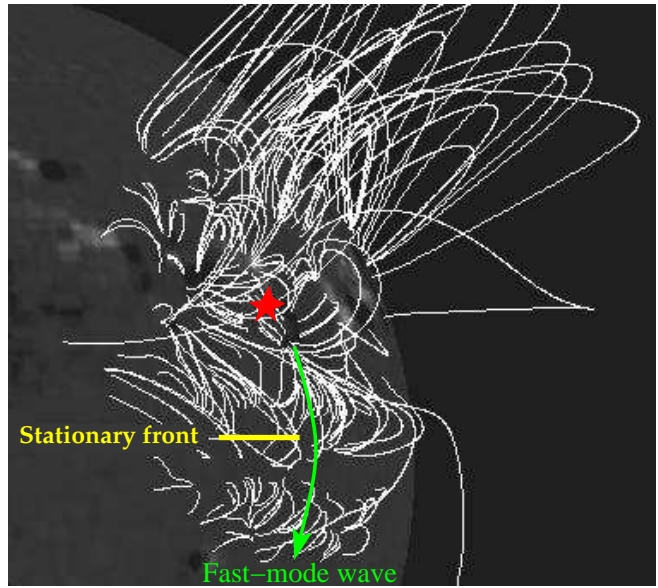} }
   \caption{A composite schematic image showing that a fast-mode MHD wave 
propagates outward ({\it green line}) from the eruption source region, passing 
through a magnetic quasi-separatrix layer (QSL) and leaving a stationary front 
({\it yellow line}) to the north of the QSL. The red star marks the eruption 
source region. The background gray-scale image is the extrapolated potential 
magnetic field around the CME/flare eruption on 2011 May 11.}
   \label{fig1}
   \end{figure}

However, recently \citet{chan16} analyzed an interesting EUV wave event, where 
they found that ahead of a slowly-moving ``EIT wave" which finally stopped at 
a magnetic QSL to form a stationary wave front, a fast-moving EUV wave passed 
through another magnetic QSL, leaving a second stationary front behind. That 
is to say, a stationary EUV wave can also be formed by the interaction between 
a fast-mode wave and a magnetic QSL. Such a stationary EUV wave front is 
different from the stationary ``EIT wave" discovered by \citet{dela99} in two 
aspects. First, in \citet{dela99}, the stationary ``EIT wave" front is 
considered as the slowly-moving ``EIT wave" asymptotically approaching the 
magnetic QSL, as simulated by \citet{chen05b, chen06b}, whereas in 
\citet{chan16}, the newly-discovered stationary EUV wave front is formed as a 
fast-mode wave passes through a magnetic QSL. Second, in \citet{dela00}, the 
stationary ``EIT wave" front is cospatial with the magnetic QSL, whereas in 
\citet{chan16}, the stationary EUV wave front deviates from the magnetic QSL, 
on the wave-incoming side, as illustrated by Figure \ref{fig1}. 

Considering that the magnetic field near a QSL is divergent and
the magnetic field (as well as the Alfv\'en speed) has a local minimum,
\citet{chan16} tentatively proposed a wave trapping model, {\it i.e.} as a
fast-mode wave propagates across a magnetic QSL which serves as a cavity,
part of the fast-mode wave is trapped inside the cavity, being reflected back 
and forth inside. Regarding the mis-alignment between the stationary wave 
front and the magnetic QSL, they suggested that it might be due to the low
accuracy of the potential field source surface (PFSS) model used to 
extrapolate the coronal magnetic field. 

In this paper, we intend to numerically simulate the interaction of a fast-mode
wave with a magnetic QSL. This paper is organized as follows. The numerical
method is described in Section \ref{sec2}, and the numerical results are
presented in Section \ref{sec3}, which is followed by discussions in Section
\ref{sec4}.

\section{Numerical Method} \label{sec2}

For the purpose of this paper, 2-dimensional (2D) ideal MHD equations are
sufficient. With the third dimension, $z$-axis, being ignored, the 2D ideal
MHD equations are shown below, which are numerically solved by the multi-step
implicit scheme \citep{hu89, chen00},

\begin{equation}
{\frac {\partial \rho}{\partial t}}+{\bf \nabla} \cdot (\rho {\bf v})=0,
\end{equation}

\begin{equation}
{\frac {\partial {\bf v}}{\partial t}}+({\bf v}\cdot {\bf \nabla}){\bf v}+
{1 \over \rho}{\bf \nabla} P -{1 \over \rho}{\bf j} \times {\bf B}+{\bf g}=0,
\end{equation}

\begin{equation}
{\frac {\partial \psi}{\partial t}}+{\bf v}\cdot {\bf \nabla} \psi=0,
\end{equation}

\begin{equation}
{\frac {\partial T}{\partial t}}+{\bf v} \cdot {\bf \nabla} T
+(\gamma-1)T{\bf \nabla} \cdot {\bf v}=0,
\end{equation}

\noindent
where the $x$-axis is horizontal, and the $y$-axis is vertical, $y=0$ 
corresponds to the solar surface, $\gamma = 5/3$ is the ratio of specific 
heats, ${\textbf g}$ is the gravity. The five independent variables are density 
($\rho$), velocity ($v_x$ and $v_y$), magnetic flux function ($\psi$), and
temperature ($T$). Here the magnetic field $\bf B$ is related to the
magnetic flux function by ${\bf B}={\bf \nabla }\times (\psi {\bf \hat{e}_z})$, 
and $\bf j={\bf \nabla }\times \bf B$ is the current density. The equations are
nondimensionalized with the following characteristic values: $\rho_0=1.67 \times 
10^{-12}$ kg m$^{-3}$, $T_0=1$ MK, $L_0=1.2\times 10^4$ km, $B_0=18.6$ G. So, the 
characteristic plasma $\beta$ is 0.02, the characteristic Alfv\'en speed is
1286 km s$^{-1}$, the Alfv\'en time scale is $\tau_A=9.33$ s.

\begin{figure}    %%%%%%%%%%%%%%%%%% FIGURE 2
   \centerline{\includegraphics[width=0.95\textwidth,clip=]{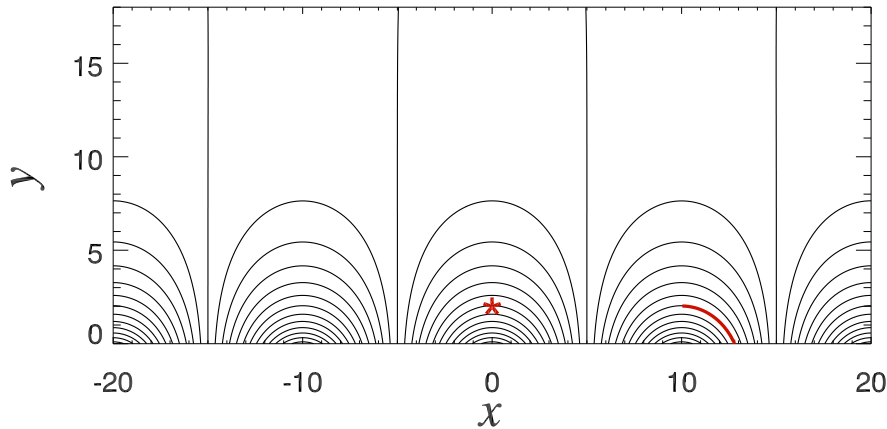} }
   \caption{Distribution of the initial magnetic field as shown by the solid lines.
The red asterisk marks the location of the pressure-enhanced area, and the red line
marks the slice used for the time-distance diagram in Figure \ref{fig8}.}
   \label{fig2}
   \end{figure}

As shown in Figure \ref{fig1}, the background magnetic field outside the
source active region in the observation is characterized by several isolated
flux systems divided by magnetic QSLs. In order to mimic such a configuration,
we generate a potential magnetic field with periodic QSLs by putting a series
of line currents with alternative directions below the solar surface, {\it i.e.}

\begin{equation}
\psi=\sum_{k=-60}^{k=60}0.5\ln[(x+20k)^2+(y+1.8)^2]-
     \sum_{k=-60}^{k=59}0.5\ln[(x+20k+10)^2+(y+1.8)^2]
\end{equation}

\noindent
in the dimensionless form. The resulting magnetic distribution is displayed in
Figure \ref{fig2}, where $x=\pm5$ and $x=\pm15$ are the locations of the
magnetic QSLs. Note that in 2D, magnetic QSLs degenerate into separatrices
\citep{demo06}.

The initial temperature is set to be uniform with the normalized $T=1$
everywhere. The initial atmosphere is in hydrostatic equilibrium, {\it i.e.} the
density decays exponentially with height in the dimensionless form 
$\rho=\exp(-\gamma gy)$. The dimensionless size of the simulation domain is
$-20\leq x \leq 20$ and $0\leq y \leq 18$. Calculation is performed in the right
half zone because of symmetry. The calculation area is discretized by $138\,
\times\,$541 grid points, which are uniformly distributed in the $y$-direction
and nonuniformly distributed in the $x$-direction, with grid points slightly
more concentrated near the $y$-axis. Symmetry conditions are specified along the
left boundary ($x=0$), whereas the top ($y=18$) and the right-hand ($x=20$) 
sides are treated as open boundaries, which allow plasmas to move out or come 
in. Besides, line-tying effect is considered on the bottom boundary, {\it i.e.} 
the values of the velocity ($v_x=v_y=0$) and $\psi$ are fixed. Besides, 
the density is also fixed, whereas the temperature gradient is set to be zero.

Strictly speaking, the fast-mode coronal shock waves, especially those that can
generate chromospheric Moreton waves, are generally thought to be driven by 
erupting CMEs, or erupting flux ropes in a strict sense, rather than by the 
pressure pulse inside solar flares \citep{cliv99, chen02}. However, in this
paper, we are not interested in the driving mechanism of the shock wave, 
therefore we use the simplest way to drive a fast-mode shock wave, {\it i.e.} by
putting an artificially high gas pressure inside a small circular area
$x^2+(y-2)^2\leq 0.5^2$, as indicated by the asterisk in Figure \ref{fig2}. 
Inside this area, the locally enhanced temperature is distributed as $T=2001-
2000[x^2+(y-2)^2]/0.5^2$, which decreases from 2001 at the center to 1 at the 
boundary of this circular area.

\begin{figure}    %%%%%%%%%%%%%%%%%% FIGURE 3
   \centerline{\includegraphics[width=0.9\textwidth,clip=]{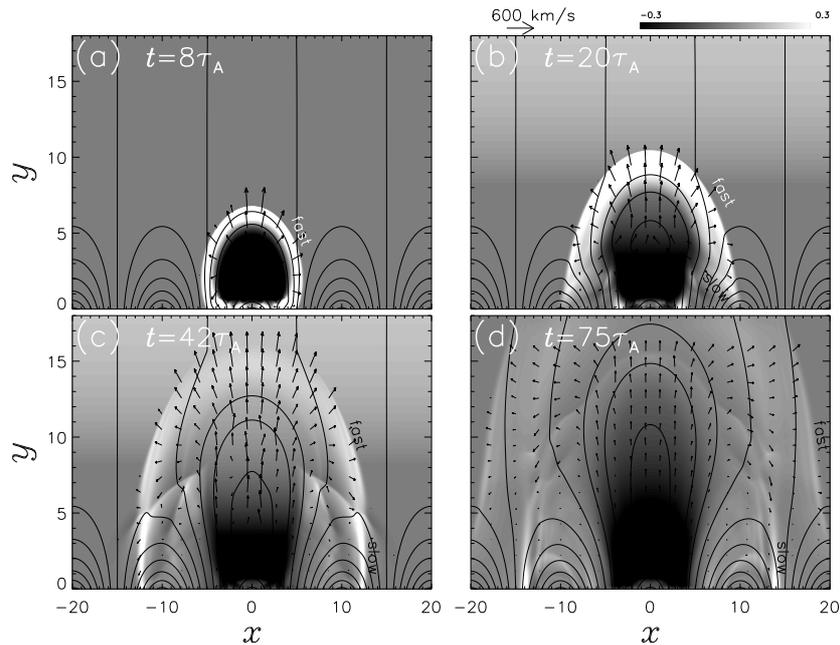} }
   \caption{Evolution of the density variation ({\it gray scale}), magnetic field 
({\it solid lines}), and velocity ({\it arrows}). Note that the dimensionless 
density enhancement near the shock wave front can be much larger than 0.3. The
scale bar is truncated at 0.3 in order to hightlight the weak variation.}
   \label{fig3}
   \end{figure}

\section{Numerical Results} \label{sec3}

The high gas pressure around the position ($x=0$, $y=2$) initiates a circular
shock wave propagating outward, as shown by the outermost wave front
which is marked by ``fast" in Figure \ref{fig3}. This figure displays the 
evolution of the base difference of the density distribution ({\it 
gray-scale}), magnetic field ({\it solid lines}), and velocity ({\it arrows}). 
As the shock wave expands, it becomes very similar to the piston-driven shock 
wave straddling over the source active region as numerically simulated by 
\citet{chen02}. The two flanks of the shock wave sweep the flux systems in the 
background and the separatrices between neighboring flux systems. As 
revealed by Figure \ref{fig3}, the shock passes through the first separatrix 
at $x=5$ (and its symmetric one on the negative $x$-axis) around $t=8\tau_A$, 
leaving nearly nothing behind. At $t=20\tau_A$, the 
footpoint of the shock wave approaches the core of the neighboring flux system. 
Since the magnetic field is stronger toward the core, the leg part of the shock 
front is refracted upward slightly. At the same time, another bright front, 
to the right of the label ``slow", appears behind the shock wave, 
which intersects with the shock wave at ($x=9.2$, $y=1.4$). At $t=42\tau_A$ as 
shown by Figure \ref{fig3}c, the bright patch, which is at around $x=12$
as marked by ``slow", becomes vertical, and actually decouples from 
the outermost shock wave. Thereafter, the outermost fast-mode shock wave keeps 
expanding rapidly, whereas the bright vertical patch moves outward extremely 
slowly. At $t=75\tau_A$ as shown by Figure \ref{fig3}d, the footpoint of the 
outermost shock wave reaches the right boundary at $x=20$. However, the bright 
patch is still at around $x=14$, as marked by the label ``slow". 
Finally, the bright patch is seen to stop near $x=14$, never reaching the 
location of the magnetic separatrix at $x=15$.

\begin{figure}    %%%%%%%%%%%%%%%%%% FIGURE 4
   \centerline{\includegraphics[width=0.9\textwidth,clip=]{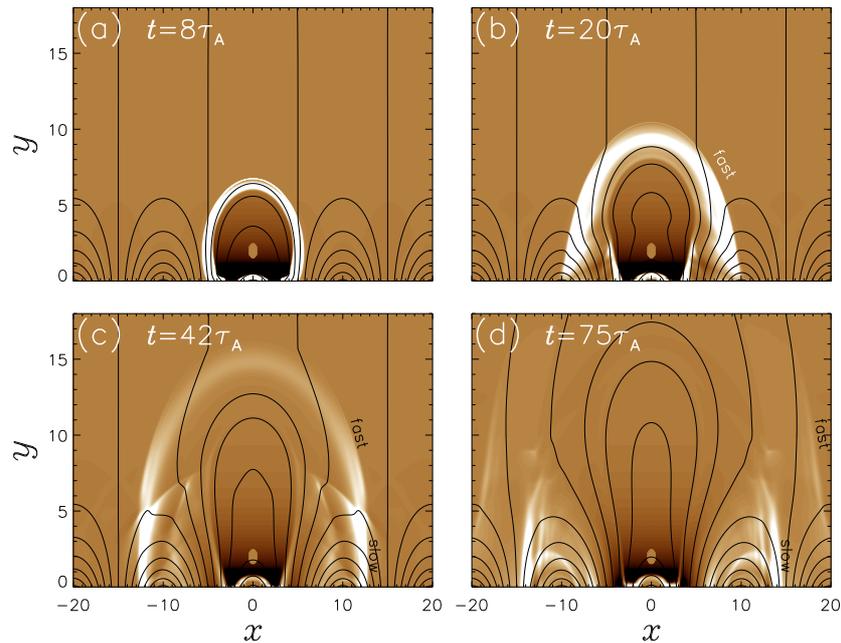} }
   \caption{Evolution of the base-difference EUV 193 \AA\ distribution ({\it color 
scale}), magnetic field ({\it solid lines}), and velocity ({\it arrows}), where the
EUV 193 \AA\ distribution is synthesized from the simulation results.}
   \label{fig4}
   \end{figure}

In order to compare the numerical results with the SDO/AIA observations, we
synthesize the AIA 193 \AA\ intensity map based on the plasma density,
temperature, and the AIA response function. The AIA 193 \AA\ emission of the
numerical results in Figure \ref{fig3} is presented in Figure \ref{fig4}, which
shows the evolution of the AIA 193 \AA\ base-difference map ({\it color}) and
the magnetic field ({\it solid lines}). Note that the base-difference map is
derived by subtracting the initial intensity from each time step. The evolution
of the 193 \AA\ map is very similar to that of the density map, which is due to
the fact that the EUV emission is proportional to the density squared.

\begin{figure}    %%%%%%%%%%%%%%%%%% FIGURE 5
   \centerline{\includegraphics[width=0.9\textwidth,clip=]{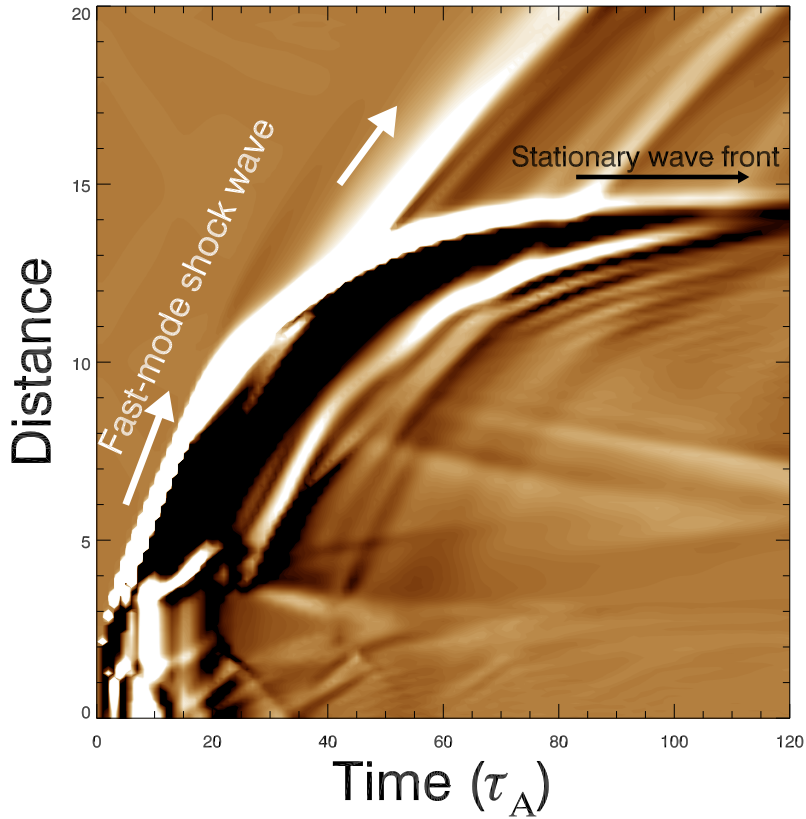} }
   \caption{Time evolution of the 193 \AA\ difference intensity distribution along
positive $x$-axis when the numerical results in Figure \ref{fig4} are viewed from
above. The start point of the distance is at $x=0$. The shock wave travels with an
initial velocity of 510 km s$^{-1}$, which decelerates to 260 km s$^{-1}$ after
$t\sim 40\tau_A$. A quasi-stationary wave front emanates from the fast-mode wave,
and approaches the distance around $x=14$.}
   \label{fig5}
   \end{figure}

Suppose that the whole evolution presented in Figure \ref{fig4} is observed 
from above, we can obtain the time-distance diagram of the wave propagation.
Figure \ref{fig5} depicts the time evolution of the AIA 193 \AA\ intensity 
distribution along the positive $x$-axis. Note that the AIA 193 \AA\ intensity
distribution is derived by integrating the 2D images in Figure \ref{fig4} over
the $y$-direction. It is seen from Figure \ref{fig5} that the fast-mode shock
wave, as indicated by the white arrow, propagates outward with an initial 
velocity of 510 km s$^{-1}$. At around $t=40\tau_A$, the shock wave front
bifurcates into two branches: A brighter one slows down rapidly, and finally
approaches but never reaches the separatrix at $x=15$ (hence we call it 
quasi-stationary wave front); The other weaker wave with a dimming edge 
{\bf (near the white thick arrow)} follows the original trajectory, and keeps 
propagating outward. The dimming edge has a velocity slightly smaller than 
510 km s$^{-1}$, but the emissive front has a velocity of only 260 km s$^{-1}$.

Figure \ref{fig5} also reveals several other wave patterns. One bright wave 
appears at the distance of 5 at $t=27\tau_A$. This wave, which is one of the 
wave train driven by the locally enhanced gas pressure, has the same fate as 
the primary shock wave, {\it e.g.} it bifurcates into a brighter 
quasi-stationary wave and a weaker fast-moving wave at $t=77\tau_A$. There is 
also a trapped wave, bouncing back and 
forth between $x=0$ and $x=2$, which results from the slow-mode wave propagating
along the magnetic loop threading the gas pressure enhanced area around 
($x=0$, $y=2$). The formation of these waves is related to the {\it ad hoc}
generation method of the primary shock wave adopted in this paper. In real 
observations, they are not necessarily present.

\section{Discussions} \label{sec4}

``EIT waves" were discovered more than 18 years ago with the SOHO/EIT telescope
\citep{thom98}. The poor cadence of the telescope, as well as the relatively 
lower spatial resolution, incurred many controversies, including the driving 
source and the nature of this spectacular phenomenon \citep[for reviews, 
see][]{will09, gall11, chen12, pats12, liu14, warm15, chen16}. Initially 
it was widely thought that ``EIT waves" are fast-mode waves. However, the 
discovery of a stationary ``EIT wave" front, as well as the significantly lower 
velocities of ``EIT waves" compared to type II radio bursts or Moreton waves, 
invoked several groups to propose non-wave models, such as the magnetic 
fieldline stretching model \citep{chen02, chen05a, chen05b, yang10, chen09}, 
the slow-mode wave model \citep{will07, wang09, mei12}, the successive 
reconnection model \citep{attr07a}, and the current shell model \citep{dela08}. 

When putting forward the magnetic fieldline stretching model for ``EIT waves", 
\citet{chen02} and \citet{chen05b} predicted that there should be two types of
waves co-existing in EUV images if only the observational cadence is high 
enough, {\it i.e.} there are in principle two EUV waves in one event, with the 
faster one being a fast-mode MHD wave and the slower one being the ``EIT waves"
due to the successive stretching of the closed magnetic field lines pushed by
an erupting flux rope.

After the launch of the SDO satellite in 2010, its onboard AIA telescope 
routinely provides EUV images with unprecedentedly high spatiotemporal 
resolutions. On the one hand, various groups reported the co-existence of two 
EUV waves in many individual events \citep{harr03, chenwu11, asai12, cheng12, 
zhen12a, shen13, whit13, xue13}. On the other hand, SDO/AIA 
revealed several features that were not seen before. For example, \citet{guo15}
found that behind the fast-mode wave (or shock wave), there is not a continual 
slower ``EIT wave". Instead, there are two or three patchy wave fronts with 
extremely low speeds such as 30 km s$^{-1}$ for each, which are more than 20 
times smaller than the speed of the fast-mode wave. However, if connected 
together, these patchy wave fronts form a pattern which is very similar to the 
classical ``EIT waves", with an apparent speed of $\sim$3 times smaller than 
that of the corresponding fast-mode wave in the same event. They illustrated how
this feature can be explained by the magnetic fieldline stretching model 
proposed by \citet{chen02}.

Another new and unexpected feature was recently found by \citet{chan16}. In
their event, besides the co-existing fast-mode wave and slower ``EIT wave" (the
latter of which finally stops at a magnetic QSL), the fast-mode wave passes
through another magnetic QSL, leaving a stationary EUV wave front behind. In
order to understand the formation mechanism of this stationary EUV wave front,
we numerically simulated the passage of a fast-mode shock wave through a
magnetic QSL. It is found that as the shock wave sweeps the isolated flux 
system, a newly-formed bright front emanates from the leg part of the fast-mode
shock wave, as indicated by Figure \ref{fig3}. This bright front is in the wake
of the fast-mode shock wave, and moves extremely slowly before finally stopping
at $x=14$, which is not cospatial but very close to the neighboring separatrix
at $x=15$. Such a quasi-stationary wave front is very similar to the 
observations analyzed by \citet{chan16}, {\it i.e.} when a fast-mode wave runs 
through an isolated flux system, a quasi-stationary EUV wave  front is left 
while the fast-mode wave keeps moving outward. More interestingly, our 
simulation results indicate that the stationary front is offset from the nearby
magnetic separatrix, which is exactly what was observed by \citet{chan16}.

\begin{figure}    %%%%%%%%%%%%%%%%%% FIGURE 6
   \centerline{\includegraphics[width=0.9\textwidth,clip=]{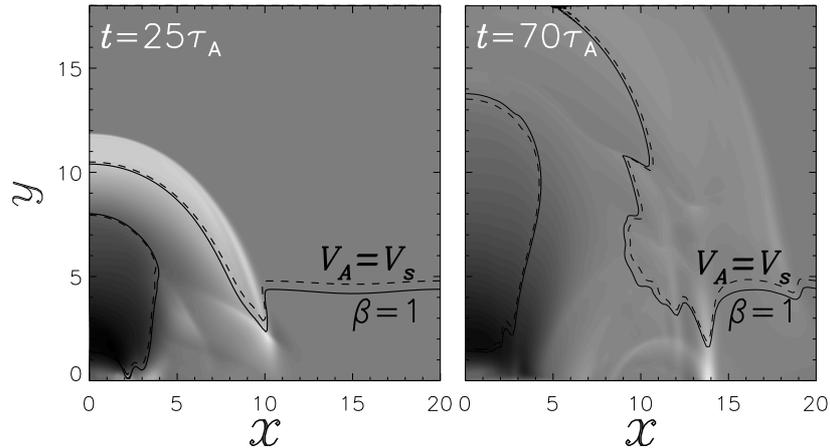} }
   \caption{Evolution of the density variation ({\it gray scale}) and the 
contour lines with the plasma $\beta=1$ and $\beta=2/\gamma$ ({\it i.e.} 
the Alfv\'en speed $V_A$ equals the sound speed $V_s$).}
   \label{fig6}
   \end{figure}

It has been well established that as a fast-mode wave penetrates into the site
with weak magnetic field where the Alfv\'en speed is comparable with the sound
speed, part of the fast-mode wave would be converted to a slow-mode wave
\citep{call05}. Such a situation can happen in the case of weak 
magnetic field, {\it e.g.} in the corona near magnetic null points
\citep{mcla06}. It can also happen in the terrestrial magnetosphere 
\citep{naka16}. In the case of an isolated magnetic flux system which is
bordered by magnetic QSLs, the coronal magnetic field above the flux system can
be very weak since the magnetic field over the QSLs are strongly divergent.
Hence, it is expected that at a certain height, the coronal Alfv\'en speed
($v_A$) is comparable with the sound speed ($v_s$), and the mode conversion can
happen here. In order to confirm this, we plot the contour lines with $\beta=1$ 
({\it solid line}) and $\beta=2/\gamma$ ({\it dashed line}) over the 
distribution of the density ({\it gray scale}) in Figure \ref{fig6}. Note that 
the plasma $\beta=2/\gamma$ corresponds to $v_A=v_s$. It is seen that around 
$t=25\tau_A$ the leg of the primary shock wave begins to bifurcate near the 
location with $\beta=1$. Whereas the ongoing fast-mode wave becomes fainter, 
the slower wave becomes remarkably bright, and moves slowly. We conjecture that
this slowly-moving bright front is a slow-mode MHD wave converted from the 
primary fast-mode wave.

\begin{figure}    %%%%%%%%%%%%%%%%%% FIGURE 7
   \centerline{\includegraphics[width=0.9\textwidth,clip=]{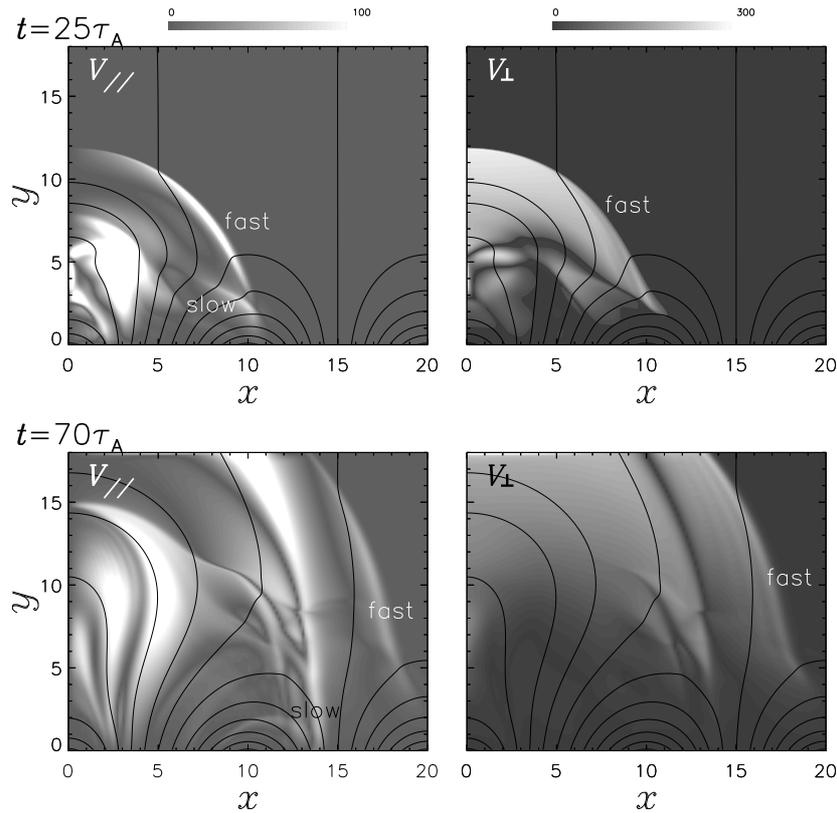} }
   \caption{Top: Distributions of ${v_\parallel}$ and ${v_\perp}$ ({\it gray 
scale}) at $t=25\tau_A$, where the magnetic field ({\it solid lines}) is superposed;
Bottom: Distributions of ${v_\parallel}$ and ${v_\perp}$ ({\it gray 
scale}) at $t=70\tau_A$, where the magnetic field ({\it solid lines}) is superposed.}
   \label{fig7}
   \end{figure}

In order to confirm the slow-mode wave nature of this bright front, we compare
the distributions of ${v_\parallel}$ and ${v_\perp}$ of the numerical results
in Figure \ref{fig7}, where ${v_\parallel}$ is the component of the plasma
velocity parallel to the local magnetic field, whereas ${v_\perp}$ is the 
component of the plasma velocity perpendicular to the local magnetic field. As
demonstrated by \citet{bogd03}, the ${v_\parallel}$ map highlights the
slow-mode waves, whereas the ${v_\perp}$ map highlights the fast-mode waves.
From Figure \ref{fig7}, it is seen that the quasi-stationary bright wave front
is indeed a slow-mode wave. It is noticed that the upper part of the primary
fast-mode shock wave is also bright in the ${v_\parallel}$ map since the upper
corona is high-$\beta$ plasma, and the fast-mode wave front there has acoustic 
nature.

\begin{figure}    %%%%%%%%%%%%%%%%%% FIGURE 8
   \centerline{\includegraphics[width=0.9\textwidth,clip=]{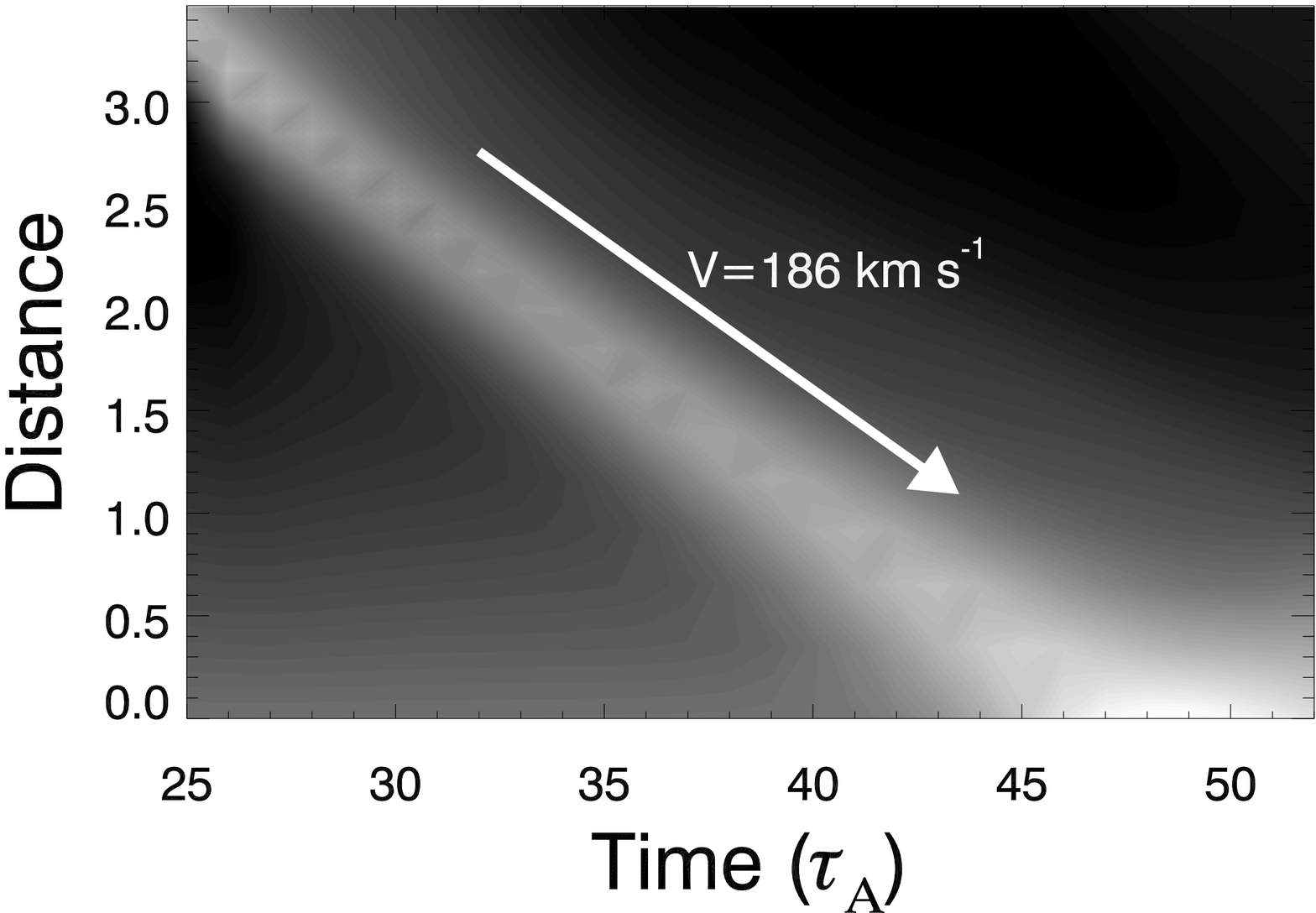} }
   \caption{Time-distance diagram of the density distribution along the slice with $\psi=7.7$ between $x=10$ and $x=12.7$. The slice is marked as the red arc in Figure
\ref{fig2}. The white ridge corresponds to the converted wave travelling along the
same magnetic field line with a propagation velocity of 186 km s$^{-1}$.  }
   \label{fig8}
   \end{figure}

The location of the bright slow-mode wave apparently moves slowly from 
$x=13$ at $t=42\tau_A$ to $x=14$ at $t=75\tau_A$ as seen from Figure 
\ref{fig4}. By examining the
movie of the evolution, it is found that each segment of the wave front is
actually  propagating along the magnetic field line. In order to derive the 
field-aligned propagation velocity, we select a curved slice along the magnetic
field line, which is marked in Figure \ref{fig2} as the thick red arc. The
time-distance diagram of the density distribution is displayed in Figure
\ref{fig8}. It is seen that the bright front travels along the closed magnetic
field line with a speed of 186 km s$^{-1}$, which is exactly the sound speed in
the simulation where the plasma temperature is $\sim$1.2 MK. This further 
confirms that the quasi-stationary bright front is a slow-mode wave converted 
from the passing fast-mode shock wave.

Having said so, another question arises: As shown by Figure \ref{fig6}, the 
$v_A=v_s$ line covers the whole layer from left to right, so why the mode 
conversion becomes evident only when the passing fast-mode wave arrives at 
$x=11$? We suggest that this is probably related to the efficiency of the mode 
conversion. According to \citet{call05}, the fast-to-slow mode conversion is the
most efficient when the wave vector is parallel to the local magnetic field. As
discerned from Figure \ref{fig3}, the mode conversion indeed happens when the
incoming wave front is nearly perpendicular to the magnetic field lines, 
{\it i.e.} the wave vector is parallel to the field lines. That is to
say, two factors lead to the wave mode conversion at $x=11$: the divergent flux
tubes near the magnetic separatrix result in a region with $\beta\sim 1$; 
meanwhile the magnetic field is roughly parallel to the shock wave normal here.

To summarize, with MHD numerical simulations, we investigated the interaction
between an incoming fast-mode shock wave and an isolated magnetic flux system
bordered by magnetic separatrices. It is revealed that when the fast-mode shock
wave penetrates into the flux system, part of the fast-mode wave is converted
to a slow-mode wave, which then travels along the magnetic field lines with the
local sound speed. Apparently the location of the wave front is shifted
across the magnetic field lines slightly, and sweeps the solar surface with a 
smaller and smaller velocity. Finally, the slow-mode wave stops at the location 
in front of the magnetic separatrix. The final location where the 
quasi-stationary front stops depends on the highest magnetic loop where the 
mode conversion occurs, since the converted slow-mode wave segment travels 
along the magnetic loop. This implies that a stationary EUV wave front, which 
was originally used as one the main reasons to challenge the fast-mode wave 
model for ``EIT waves" \citep{dela99, chen05b}, can also be produced by 
fast-mode waves via the mode conversion at the layer where the Alfv\'en speed 
is comparable to the sound speed. The slow-mode wave, trapped inside the 
magnetic loop, travels along the magnetic field line to the footpoint of the
field line, forming a quasi-stationary wave front. However, it should be pointed
again that such a stationary wave front is different from the stationary ``EIT
waves" in two aspects: (1) In the former case, the incident fast-mode wave keeps
going, leaving a stationary wave front behind. However, in the latter case, the
slowly-moving ``EIT wave" gradually decelerates to form a stationary front;
(2) The stationary wave front generated by the mode conversion from a passing
fast-mode wave is offset from the magnetic QSL, whereas the stationary ``EIT 
wave" is generally cospatial with the magnetic QSL, as demonstrated by
\citet{dela00} and \citet{chen05b, chen06b}. Future spectroscopic 
observations as in \citet{madj15} and \citet{vann15} might dig out more
differences between these two types of stationary fronts. 

It is noted in passing that fast-mode MHD waves are frequently generated by the
CME/flare eruptions \citep{asch99, naka99, liu11, su15, vrsn16}, and magnetic 
QSLs are also distributed all over the solar surface, therefore, the mode 
conversion studied in this paper might happen frequently, which deserves further
studies.
%%%%%%%%%%%%%%%%%%%%%%%%%%%%%%%%%%%%%%%%%%%%%%%%%%%%%%%%%%%%%%%%%%%%%%%%%%%
%% Acknowledgements
%
\begin{acks}
The authors are grateful to the referee for the constructive suggestions. 
This research was supported by the Chinese foundations NSFC (11533005 and 
11025314), 2011CB811402, and Jiangsu 333 Project. PFC thanks the ISSI-Beijing
office for organizing a wave workshop, where the discussions inspired 
the ideas presented in this paper.
\end{acks}

%%% %%%%%%%%%%%%%%%%%%%%%%%%%%%%%%%%%%%%%%%%%%%%%%%%%%%%%%%%%%%
\bibliographystyle{spr-mp-sola-cnd}
\bibliography{ref}

%\end{barticle}
%\endbibitem

%\end{thebibliography}
%%%%%%%%%%%%%%%%%%%%%%%%%%%%%%%%%%%%%%%%%%%%%%%%%%%%%%%%%%%%%%%%%%%%%%%%%%%
\end{article}
\end{document}